# Calculation of the Tafel slope and reaction order of the oxygen evolution reaction between pH 12 and pH 14 for the adsorbate mechanism

Denis Antipin, Marcel Risch*

Nachwuchsgruppe Gestaltung des Sauerstoffentwicklungsmechanismus, Helmholtz-Zentrum Berlin für Materialien und Energie GmbH, Hahn-Meitner Platz 1, 14109 Berlin, Germany

* marcel.risch@helmholtz-berlin.de

**Abstract**

Despite numerous experimental and theoretical studies devoted to the oxygen evolution reaction, the mechanism of the OER on transition metal oxides remains controversial. This is in part owed to the ambiguity of electrochemical parameters of the mechanism such as the Tafel slope and reaction orders. We took the most commonly assumed adsorbate mechanism and calculated the Tafel slopes and reaction orders with respect to pH based on microkinetic analysis. We demonstrate that number of possible Tafel slopes strongly depends on a number of preceding steps and surface coverage. Furthermore, the Tafel slope becomes pH dependent when the coverage of intermediates changes with pH. These insights complicate the identification of a rate-limiting step by a single Tafel slope at a single pH. Yet, simulations of reaction orders complementary to Tafel slopes can solve some ambiguities to distinguish between possible rate-limiting steps. The most insightful information can be obtained from the low overpotential region of the Tafel plot. The simulations in this work provide clear guidelines to experimentalists for the identification of the limiting steps in the adsorbate mechanism using the observed values of the Tafel slope and reaction order in pH-dependent studies.

**Introduction**

Electrochemical water splitting is a promising approach to store excess energy in the form of hydrogen bonds [1–3]. It comprises the hydrogen evolution reaction ($4H_2O + 4e^- \rightarrow 2H_2 + 4OH^-$; in alkaline media) and oxygen evolution reaction ($4OH^- \rightarrow O_2 + 2H_2O + 4e^-$, also in alkaline media). Latter reaction is known as a bottleneck in total efficiency due to it sluggish kinetics [4–6]. Despite numerous studies devoted to the OER, the mechanism of this reaction remains controversial due to lack of direct experimental evidence. This is mainly caused by the fact that oxygen evolution reaction is a complex process that requires the transfer of four electrons and four protons. Most of our knowledge about likely mechanisms is obtained from theoretical works, either by density functional theory or kinetic modeling or a combination of both [7–18]. The pH dependence of the calculated properties has thus far received little attention.

Insight into the mechanism of OER can be gained by evaluating the Nernst slope $\nu = (\partial E/\partial pH)_{\log i}$, the Tafel slope $b = (\partial E/\partial \log i)_{pH}$, and the reaction order with respect to pH $\rho = (\partial \log i/\partial pH)_E$ that are related as [19,20]:

$$\left(\frac{\partial E}{\partial pH}\right)_{\log i} = -\left(\frac{\partial E}{\partial \log i}\right)_{pH} \times \left(\frac{\partial \log i}{\partial pH}\right)_E \quad \text{(Eq. 1)}$$

These partial derivatives are the key parameters of electrochemical reaction kinetics. Their values may depend on the overpotential and pH. Most commonly, the mechanistic analysis is focused on the evaluation of the Tafel slope. It is applied to the Faradaic region of the obtained voltammograms or to chronoamperometry, where the process of interest (OER in our case) occurs. Later, the observed Tafel slopes can be compared to simulated or calculated ones to elucidate differences in the mechanism [20,21]. However, different observed Tafel slopes for similar materials or the same slope for different catalysts are insufficient to uniquely assign the rate-limiting step and the mechanistic sequence [20,22,23].

In this work we calculated two key mechanistic parameters of OER by microkinetic analysis, namely the Tafel slope and reaction order with respect to pH. They were calculated specifically for the adsorbate mechanism [11,12,24]. The calculated values and observable Tafel slopes strongly depended on the choice of the resting state and on the kinetic constants as well as on pH in some cases. Finally, we identify trends in the simulated data to provide guidelines for the assignment of the most likely rate-liming step based on the Tafel slope and reaction order.

**Methods**

The mechanistic parameters of Tafel slope and reaction order were calculated by microkinetic analysis [25]. In this work, it was based on basic electrochemical equations and expressions for current density, reaction rate and kinetic constants.

Current density of electrochemical reaction can be written as

$$j = I/A = n \cdot F \cdot r; \quad \text{(Eq. 2)}$$

where n – number of electrons involved; F – Faradaic constant (= 96500 C/mol); A – surface area of the catalyst; r – reaction rate.

The reaction rate can be described by the following equation:

$$r = k_{+i/-i} \cdot a_j \cdot \theta_k; \quad \text{(Eq. 3)}$$

where $k_{i/-i}$ – kinetic constant of the direct/reverse reaction; $a_j$ – activity of involved species (OH, $H_2O$ or $O_2$); $\theta_k$ – surface coverage.

For all simulations, the activity of water in aqueous solution ($a_{H2O}$) was equal to 1 and activity of dissolved oxygen ($a_{O2}$) was assumed to be 0.001 [26]. Specifically, for Tafel analysis a hydroxide activity of $a_{OH}$ = 1 was used, corresponding to pH = 14. For analysis of the reaction order, the hydroxide activity ranged from 0.01 to 1 (corresponding to pH 12 -14) to stay in the alkaline regime.

The expression for the kinetic constant depends on the nature of the step. For electrochemical steps,

$$k_{+i} = k_{+i}^0 \cdot e^{(1-\alpha) \cdot f \cdot \eta} \text{ - for the forward reaction;} \quad \text{(Eq. 4)}$$

$$k_{-i} = k^0_{-i} \cdot e^{-\alpha \cdot f \cdot \eta} \text{ - for the backward reaction,} \tag{Eq. 5}$$

where $\alpha$ – transfer coefficient; $f = RT/F$; R – universal gas constant (= 8.314 J mol$^{-1}$ K); T – temperature in K; F – Faraday constant; η = E$_{appl}$ - E$^0$ -overpotential, where E$^0$ is the thermodynamic equilibrium potential [27]. Here, we use E$^0$ = 0.40 V (standard potential of the OER at pH 14).

The expression for chemical steps is

$$k_{+i/-i} = k^0_{+i/-i}. \tag{Eq. 6}$$

In our analysis, the transfer coefficient $\alpha$ was set equal to 0.5, unless otherwise stated. We also assumed clearly greater backward kinetic constants as compared to the forward ones ($k^0_{-i} \gg k^0_i$). With this assumption, all steps prior to RLS do not happen simultaneously and all possible values for mechanistic parameters can be observed.

Combining Eqs. 2-6, we formulate the final expressions for the current density:

$$j = n \cdot F \cdot k^0_{+i} \cdot e^{(1-\alpha) \cdot f \cdot \eta} \cdot a_j \cdot \theta_k \text{ – for forward electrochemical step;} \tag{Eq. 7}$$

$$j = n \cdot F \cdot k^0_{-i} \cdot e^{-\alpha \cdot f \cdot \eta} \cdot a_j \cdot \theta_k \text{ – for backward electrochemical step;} \tag{Eq. 8}$$

$$j = n \cdot F \cdot k^0_{+i/-i} \cdot a_j \cdot \theta_k \text{ – for chemical steps.} \tag{Eq. 9}$$

We assumed surface coverage of the OH intermediate equal to $\theta_{OH}$ = 1 for the first step in the mechanism, meaning that the whole surface is fully covered with just one species. If there are some other species on the surface, in other words $\theta_{OH} \neq 1$, it will affect only current density values, but not the observed Tafel slopes or reaction order. When further steps were considered to be rate-limiting, expressions for surface coverages were obtained through an assumption of pre-equilibrated previous steps. Thus, for all reactions prior to RLS, the reaction rates of the forward and backward steps are equal ($r_i = r_{-i}$). Considering that the sum of all possible surface coverages is equal to 1, we derived systems of equations for the surface coverage of all intermediates $\theta_k$. Solving these systems, expressions for all surface coverages were obtained and put in Eqs. 7 or 9 for the rate-limiting step. An example calculation for the first two steps can be found in the Supporting Information.

**Results and discussion**

*Introduction of the treated mechanism*
We calculated Tafel slope and reaction order with respect to pH by the microkinetic procedure outlined in the previous section and briefly summarized in Fig. 1. While Tafel slopes are often calculated assuming limiting a low or high coverage limit and further simplifications [21,22,28], our approach explicitly includes the transitions between the limiting states in coverage that are also observable experimentally, e.g., as non-constant Tafel slopes in a given range of current or overpotential.

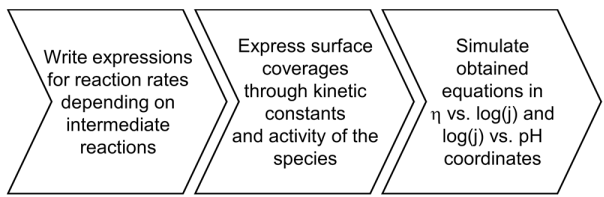

**Fig. 1.** Scheme of the simulation procedure with key steps.

Prototypical mechanisms of the OER have recently been proposed by us [29]. We focus in this report on the "adsorbate evolving mechanism (AEM)" or simply adsorbate mechanism, of which an early variant was proposed by Kobussen et al. [30]. The mechanism is popular in more recent DFT work [7,11] and frequently assumed in experimental studies, e.g., in refs. [31,32]. It is also assumed for the reverse reaction of oxygen reduction [33,34]. The proposed steps in alkaline electrolytes are:

*M1*     M-OH + OH$^-$ ⇌ M=O + H$_2$O + e

*M2*     M=O + OH$^-$ ⇌ M-OOH + e

*M3*     M-OOH + OH$^-$ ⇌ M-OO$^-$ + H$_2$O

*M4*     M-OO$^-$ ⇌ M + O$_2$ + e

*M5*     M + OH$^-$ ⇌ M-OH + e

We here assumed that the catalysts have a hydroxylated surface in their resting state in alkaline electrolytes, which is supported experimentally by ambient pressure XPS studies [35–37]. Our sequence is also used in experimental work by other groups [38–40]. Yet, it is in contradiction to other theoretical work [7,10,41], where the surface is assumed to be bare (not hydroxylated), i.e., the sequence starts with *M5* in our notation. This is a very important point as the choice of the resting state affects the calculated mechanistic parameters (Fig. S1 and Table S1). Here, all mechanistic parameters are calculated based on *M1 - M5*. These calculations rely on the coverage of surface states that are underlined in our notation.

**Table 1.** Values of kinetic constants used in the simulations as discussed in the text.

| Kinetic constant | Rate-limiting step | | | | | |
|---|---|---|---|---|---|---|
| | *M1* | *M2* | *M3* | *M3'* | *M4* | *M5* |
| $k_{+1}$ | 1 | $10^3$ | $10^3$ | $10^3$ | $10^3$ | $10^3$ |
| $k_{-1}$ | – | $10^7$ | $10^7$ | $10^7$ | $10^7$ | $10^7$ |
| $k_{+2}$ | – | 1 | $10^3$ | $10^3$ | $10^3$ | $10^3$ |
| $k_{-2}$ | – | – | $10^{10}$ | $10^6$ | $10^{10}$ | $10^{10}$ |
| $k_{+3}$ | – | – | 1 | 1 | $2\cdot10^{12}$ | $2\cdot10^{12}$ |
| $k_{-3}$ | – | – | – | – | $10^{12}$ | $10^{12}$ |
| $k_{+4}$ | – | – | – | – | 1 | 10 |
| $k_{-4}$ | – | – | – | – | – | $10^{12}$ |
| $k_{+5}$ | – | – | – | – | – | 1 |
| $k_{-5}$ | – | – | – | – | – | – |

Our simulations are based on the kinetic constants found in Table 1. They are consistently and systematically chosen but arbitrary as very little experimental measurements are available [42–45]. The rate limiting step always has a rate of 1 (for simplicity) and all other steps are orders of magnitude faster. While the relative values are similar to experiments [42], the absolute values in Table 1 are consequently orders of magnitude higher. Thus, we use arbitrary units for the current density in all figures. The aim of this report is uncovering possible trends rather than an accurate calculation of the parameters for a given catalyst.

*Simulation of the Tafel slope*

There was a clear dependence of Tafel slope on surface coverage in our calculations. Fig. 2 shows simulated Tafel plots assuming the indicated steps *M1* to *M5* as rate-limiting. We first considered the deprotonation of an OH-covered surface as limiting (*M1*), for which we found a single Tafel slope of $b_{M1}$ = 118 mV·dec$^{-1}$. The number of possible different Tafel slopes increased with a number of preceding steps. Moving to the (*M2*) reaction, we observed the two slopes of 40 mV·dec$^{-1}$ and 118 mV·dec$^{-1}$, depending on the dominant surface coverage of either M=O (product of *M1*) or M-OOH (product of *M2*). This means that the Tafel slopes depend on the pre-equilibria of preceding steps as also discussed elsewhere in the case of limiting coverages [7,10]. When the coverage of surface states was not constant, the Tafel slopes had intermediate values that strongly depend on the applied overpotential or current. Thus, a transition between two constant Tafel slopes in an experiment indicated that the dominant surface coverage changed. The overpotential at which the transition was observed depended on the choice of the kinetic constants (Fig. 2c vs 2d) as discussed further below. The third step (*M3*) in our sequence was chemical, which resulted in an infinite Tafel slope $b_{M3}$ = ∞ mV·dec$^{-1}$ at high overpotentials (Fig. 2c). Infinite Tafel slopes are not found experimentally, however values above 200 mV·dec$^{-1}$ can be found in literature for Cu meshes or Pt plates [46,47]. The latter steps *M4* and *M5* produced complicated plots with many Tafel slopes and transitions (Fig. 2e,f). However, these cases are experimentally less relevant as usually, early RLS are assumed and supported by experimental data of electrocatalysts [42,48,49]. In particular, the expected Tafel slopes at low overpotential below about 30 mV·dec$^{-1}$ are rarely reported (Fig. S2a and recent reviews in refs. [50–53]).

The observation of multiple Tafel slopes depended strongly on the choice of the kinetic constants. We illustrated this for the case where *M3* is the RLS (Fig. 2c,d). For this step, we used the same values for kinetic constants except $k_{-2}$. It is $10^5$-fold lower for *M3'* (Table 1) and resulted in lower values of $\theta_O$, a narrowing of the coverage distribution as function of overpotential and a shift of the maximal coverage to a different overpotential (here a lower one; blue curves in Fig. 2c,d). The latter choice of kinetic constants means that M=O transforms into M-OOH much faster so that a surface fully covered with the M=O surface state cannot be observed. In other words, the overpotential range, for which a Tafel slope of $b_{M3}$ = 59 mV·dec$^{-1}$ (O-O bond formation from M=O to M-OOH) is expected becomes so narrow that it cannot be resolved. The same arguments can be made for all other limiting steps in the mechanism. In general, if the kinetic constants of the forward reaction ($k_{+i}$) were faster than that of the backward reaction ($k_{-i}$) then the changes in surface state were negligible and our plot reduced to the high overpotential/current limit that is either 118 mV/dec (assuming α = 0.5) for limitation by an electrochemical and infinite for a chemical step (Fig. S3 and Table S2).

This scenario was proposed based on experiments for NiFe LDH [42] but not for Sb-doped SnO$_2$, PbO$_2$ and boron doped diamond, arguably much worse catalysts for the OER [43]. Thus, we show and discuss the more general result with $k_{+1} < k_{-1}$ with the caveat that not all Tafel slopes calculated herein may be experimentally observed on all electrocatalysts.

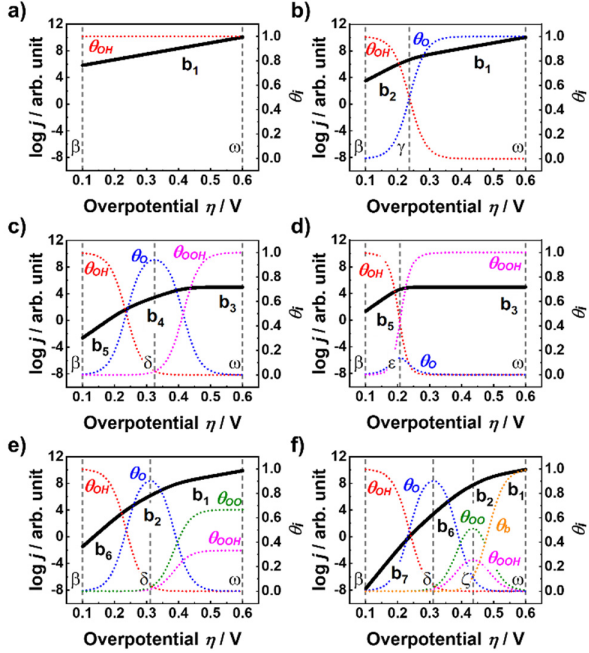

**Fig. 2.** Simulated Tafel plots assuming the RLS of (a) *M1*, (b) *M2*, (c) *M3*, (d) *M3'*, (e) *M4* and (f) *M5*. *M3* and *M3'* have the same RLS but were simulated with different sets of kinetic constants found in Table 1. All simulations use a transfer coefficient of $\alpha = 0.5$. The obtained Tafel slopes are denoted as: $b_1 = 118$ mV·dec$^{-1}$, $b_2 = 40$ mV·dec$^{-1}$, $b_3 = \infty$ mV·dec$^{-1}$, $b_4 = 60$ mV·dec$^{-1}$, $b_5 = 30$ mV·dec$^{-1}$, $b_6 = 24$ mV·dec$^{-1}$, $b_7 = 17$ mV·dec$^{-1}$. $\beta$ to $\omega$ denote potentials of interest that are analyzed further in other figures. Dataset in ref. [54].

### *Effect of the transfer coefficient on the Tafel slope*

Another important parameter that affects Tafel slope is the transfer coefficient. In general, the anodic transfer coefficient is defined as [55]

$$\alpha = \frac{RT}{F} \times \left(\frac{dlnj_a}{dE}\right). \tag{Eq. 10}$$

The OER is an entirely anodic reaction with high overpotential, which allows neglecting the contribution of the cathodic transfer coefficient. By definition, $\alpha$ is linked with the Tafel slope, which is defined as mentioned above as

$$b = \left(\frac{\partial E}{\partial \log i}\right)_{pH} = \frac{RT}{\alpha F}. \tag{Eq. 11}$$

Whereas the transfer coefficient can be easily determined for an electrode reaction consisting of a single one electron transfer step, it becomes controversial for complex electrode reactions such as the OER. Currently the most common approach in kinetic investigations of electrode processes is to ascribe $\alpha$ an arbitrary value of 0.5, both theoretical [7,10] and experimental [7,56] studies, which results in associated Tafel slopes of 118, 59, 40 mV·dec$^{-1}$ and lower ones (Table 2, $\alpha$ = 0.5). However, recent studies suggest that the value of $\alpha$ can be higher ~ 0.8-0.9 [57,58]. Additionally, Tafel slopes that cannot be explained by $\alpha$ = 0.5 such as 80, 90 and >120 mV·dec$^{-1}$ have been reported in literature (Fig. S2b), which supports that $\alpha$ = 0.5 is only a special case among other possible values. This literature survey motivated us to simulate Tafel plots with different values of the transfer coefficient (Fig. 3). Here we used moderate values of $\alpha$ (0.25, 0.5 and 0.75; schematic reaction diagram in Fig. 3a) to illustrate the expected trends in the Tafel plots.

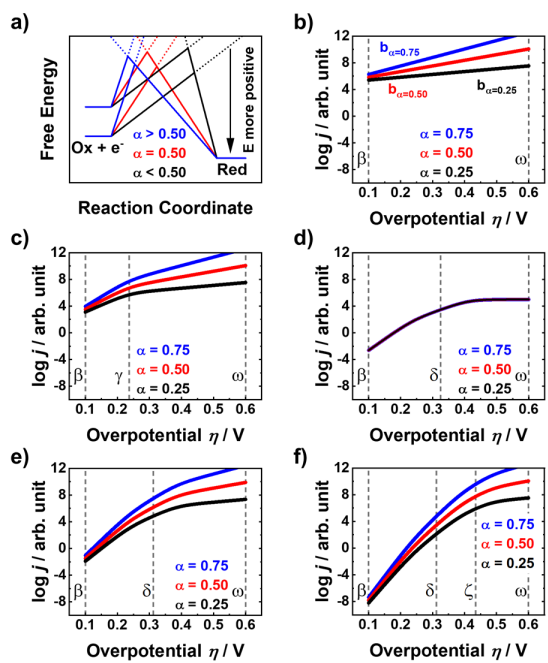

**Fig. 3.** Free energy diagram of Ox + e = Red reaction with different transfer coefficient values (a) and simulated voltammogram plots with different values of $\alpha$ assuming the RLS of (b) M1, (c) M2, (d) M3, (e) M4 and (f) M5. Blue, red and black lines correspond to $\alpha$ values of 0.75, 0.5 and 0.25 respectively. All obtained values of the Tafel slope can be found in Table 2. Dataset in ref. [54]

The plots in Fig. 3 are qualitatively similar to those in Fig. 2. Three important observations were made. Firstly, the transition points between overpotential ranges of fixed Tafel slope were not affected by the choice of $\alpha$. Secondly, the values of the Tafel slopes were only affected by $\alpha$ for the electrochemical rate-limiting steps (Fig. 3b,c,e,f) but not for chemical rate-limiting states (Fig. 3d). It is no surprise since $\alpha$ by definition describes electrochemical charge transfer and is thus connected with current and potential. It can also be clearly seen in equations for electrochemical RLS (Eqs. 7,8) and absence of $\alpha$ for chemical RLS (Eq. 9). Thirdly, the value of the constant Tafel slopes clearly depended on the choice of $\alpha$, where $\alpha$ < 0.5 increased the Tafel slope and $\alpha$ > 0.5 decreased the Tafel slope. The largest difference can be observed for the highest value of the Tafel slope, i.e., 79 mV·dec$^{-1}$ for $\alpha$ = 0.75 and 236 mV·dec$^{-1}$ for $\alpha$ = 0.25. For better numerical demonstration, we calculated all possible Tafel slopes using a common formula for multiple-electron reactions that connects Tafel slope, transfer coefficient and number of electrons involved (at 25 °C) [21]:

$$b = \left(\frac{\partial E}{\partial \log i}\right)_{pH} = \frac{2.303RT}{F} \times \frac{1}{n_b + \alpha n_d} \approx \frac{59 \text{ mV} \cdot \text{dec}^{-1}}{n_b + \alpha n_d} \qquad \text{(Eq. 12)}$$

All obtained values are summarized in Table 2, whose entries are color coded according to three categories: (i) improbable reaction steps in red, i.e., requiring more than 1 electron to be involved during the RLS; (ii) probable electrochemical RLS in blue; and (iii) probable chemical RLS in green. It can be clearly seen in Fig. 3 and Table 2 that the difference in Tafel slope with $\alpha$ decreased for smaller values of the Tafel slope, for more electrons were transferred before the RLS. Though they are distinct in the theoretical calculations, experimental verification might be impeded due to noise, small ranges with constant Tafel slope and other factors. Experimental errors of the Tafel slope can be as low as 1-2 mV dec$^{-1}$ [20] (one standard deviation, 68 % confidence interval) or 3-5 mV dec$^{-1}$ [59] (two standard deviations, 95 % confidence interval) for independent measurements of identical prepared electrodes. This indicates that the small deviations with $\alpha$ for more than n = 2 are unlikely to be resolved experimentally in typical experimental studies. Our survey of reported Tafel slopes further supported the cutoff criterium of 2 transferred electrons before the RLS because Tafel slopes below 30 mV dec$^{-1}$ are rarely reported, while our literature survey showed clear maxima near 118, 79, 59 and 40 mV dec$^{-1}$ (Fig. S2), supporting a ≥ 0.5.

**Table 2.** Variation of the Tafel slope with transfer coefficient $\alpha$.

| Step | n | $n_b$ | $n_d$ | Tafel slope | | |
|---|---|---|---|---|---|---|
| | | | | $\alpha$ = 0.25 | $\alpha$ = 0.5 | $\alpha$ = 0.75 |
| M1 | 1 | 0 | 1 | 236 | 118 | 79 |
| M1 | 1 | 1 | 0 | 59 | 59 | 59 |
| M2 | 2 | 0 | 2 | 118 | 59 | 40 |
| M2 | 2 | 1 | 1 | 48 | 40 | 34 |
| M2 | 2 | 2 | 0 | 30 | 30 | 30 |
| M3 | 3 | 0 | 3 | 79 | 40 | 27 |
| M3 | 3 | 1 | 2 | 40 | 30 | 24 |
| M3 | 3 | 2 | 1 | 27 | 24 | 22 |
| M3 | 3 | 3 | 0 | 20 | 20 | 20 |
| M4 | 4 | 0 | 4 | 59 | 30 | 20 |
| M4 | 4 | 1 | 3 | 34 | 24 | 18 |
| M4 | 4 | 2 | 2 | 24 | 20 | 17 |
| M4 | 4 | 3 | 1 | 18 | 17 | 16 |
| M4 | 4 | 4 | 0 | 15 | 15 | 15 |

n denotes total number of the electrons involved in the step, $n_b$ – number of electrons before the RLS, $n_d$ – number of electrons during the RLS. Red rows stand for highly improbable reactions, blue row – for electrochemical RLS, green row – for chemical RLS. A double bar indicates the boundary of experimental verification (see text).

In summary, there were several general trends in the Tafel slopes that must be highlighted. First, the lowest possible Tafel slope for the chosen RLS occured at the lowest overpotential/current region. It might cause some problems in experimental verification since catalysts with low overpotentials are required, at which catalytic currents may overlap with metal redox peaks. Second, all electrochemical steps approached 118 mV·dec$^{-1}$ slope, while all chemical steps approached an infinite slope as expected from the limiting cases [21]. These slopes can be accessed at high overpotential/current region and thus, can be obtained experimentally, often without interference by other redox peaks. These trends are also seen in the earlier microkinetic work of Shinagawa et al. [10] on the adsorbate mechanism, which is based on different assumptions of kinetic constants and a non-hydroxylated resting state. While the earlier work gives qualitatively similar trends with overpotential, the values of the constant Tafel slopes in the plots for each RLS (obtained only at $\alpha$ = 0.5) differ from our work. Lastly, the value of the Tafel slope depended on the transfer coefficient for electrochemical rate-limiting steps but did not affect the transition between regions of constant Tafel slope or Tafel slopes with chemical rate-limiting steps. The calculated differences in Tafel slope with the transfer coefficient may be too small to be resolved experimentally for more than two transferred electrons.

*Effect of pH on the Tafel slope*

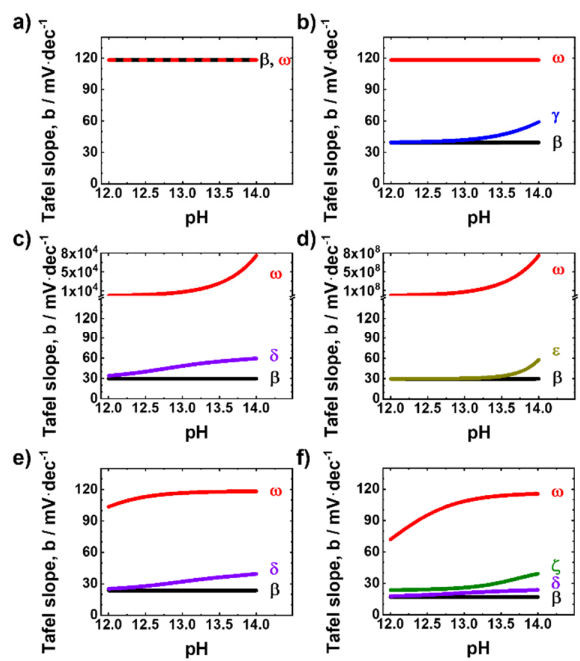

**Fig. 4.** Simulated Tafel slopes as function of pH at key potentials identified in Fig. 2 assuming the RLS of (a) *M1*, (b) *M2*, (c) *M3*, (d) *M3'*, (e) *M4* and (f) *M5*. *M3* and *M3'* have the same RLS but were simulated with different sets of kinetic constants found in Table 1. All simulations use a transfer coefficient of $\alpha$ = 0.5. Dataset in ref. [54].

The Tafel slope may depend on the pH (Fig. 4). To investigate this point, we simulated the pH dependence of the Tafel slope for selected overpotentials labeled $\beta$ to $\omega$. (Note that $\alpha$ is reserved for the transfer coefficient in this manuscript) Their values may be found in Fig 2 and 3. There was no change in coverage when *M1* was the RLS and the Tafel slope was 118 mVdec$^1$ between pH 12 and 14 (Fig. 4a). When *M2* was the RLS, $\beta$ and $\omega$ remained independent of pH but had different values (Fig. 4b). The Tafel slope at potential $\gamma$ was nearly constant for low pH but became pH dependent above pH 13. At potential $\delta$, the Tafel slope changed for each investigated pH with a sigmoidal shape, which could be seen when *M3*, *M4* or *M5* were the RLS (Fig. 4c,e,f). The trends of overpotentials $\varepsilon$ and $\xi$ matched those of potential $\delta$, i.e., constant at low pH and pH-dependent for high pH.

The pH trends in the Tafel slope can be rationalized by a change in the population of the intermediates. We considered *M3* as the RLS (Fig. 4c,d) to illustrate the effect of the surface population of intermediates on the pH-dependence of the Tafel slope. At potential $\delta$, the coverage with intermediate M-O was at a maximum ($\theta_O$ = 1.0), while the coverage of intermediates M-OH and M-OOH were at a minimum at pH 14 (Fig. S4). Reducing the pH to pH 12 lowered the population of M-O to $\theta_O$ = 0.2. As the Tafel slope depends on the surface coverage, it became pH-dependent through the change in the population of the intermediates with pH. The comparison between Fig. 4c and 4d illustrates that the pH trend fundamentally depended on the kinetic constants, which determined whether the coverage and Tafel slope changed in a given pH range (e.g., Fig. 5c or pH >13 in Fig. 4d) or did not change (e.g., pH <13 in Fig. 4d).

Even though analysis of the Tafel slope gives useful insights into mechanism of oxygen evolution reaction, it is not sufficient to uniquely ascertain the RLS. For instance, Tafel slope of 40 mV·dec$^{-1}$ can be observed for three different steps as RLS, namely *M2*, *M4* and *M5* (Fig. 2b, e, f). In order to distinguish between these different possible RLS, further orthogonal mechanistic parameters need to be obtained. Tafel analysis can be performed as function of pH (i.e., a range of OH$^-$ activities) to obtain the reaction order with respect to pH or the reactant (OH$^-$ in alkaline) from the slopes of log *j* vs. pH to gain complementary mechanistic insight.

*Simulation of the reaction order with respect to pH*

We used microkinetic analysis to obtain the reaction order with respect to the overpotential on the scale of the standard hydrogen electrode (SHE) [19] because its value trends with the number of transferred hydroxide similar to the number of transferred electrons obtained from Tafel analysis. The reaction order is obtained as the derivative of the logarithmic current with respect to pH (Fig. S5), similarly to the derivatives that yield the simulated Tafel slopes above. Akin to Tafel slope simulations, we made an analysis of the reaction order for different transfer coefficient values for two extreme steps – *M1* (Fig. S6a) and *M5* (Fig. S6b). As it was mentioned above, the Tafel slope of electrochemical rate-limiting steps up to n = 2 is strongly connected with the transfer coefficient. In the plot of log j vs. pH, a change in the transfer coefficient affected the y-axis intercept of the resulting trend of the current with pH, but did not affect the slope regardless of the RLS. Thus, reaction order with respect to pH may be a perfect match to the Tafel slope that helps to mitigate the ambiguity introduced by the uncertainty in the transfer coefficient.

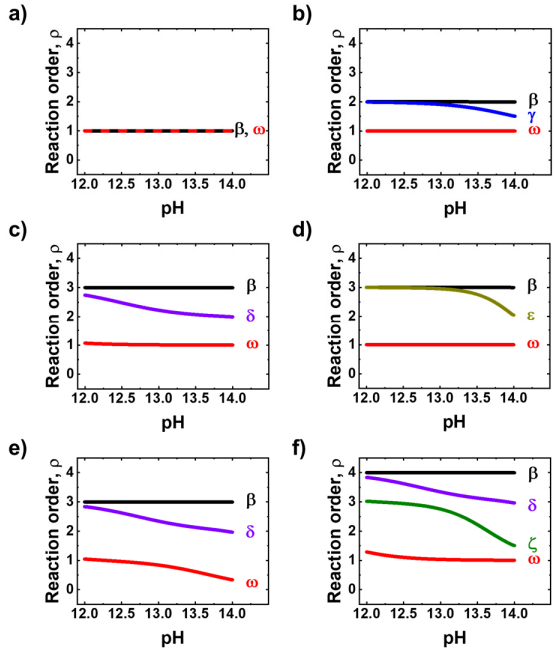

**Fig. 5.** Simulated reaction order (ρ) plots assuming for selected key potentials in Fig. 2 assuming the RLS of (a) *M1*, (b) *M2*, (c) *M3* (d) *M3'*, (e) *M4* (e) and (f) *M5* steps. *M3* and *M3'* have the same RLS but were simulated with different sets of kinetic constants found in Table 1. Dataset in ref. [54]

Figure 5 shows the expected reaction orders for the potentials of interest defined in Fig. 2. We assume that steps *M1-M5* were the RLS akin to the Tafel analysis above. Again, we considered deprotonation of the M-OH surface state (*M1*) as the first possible RLS. Analogous to the Tafel slope, we observed only one possible reaction order $\rho_{M1}$ = 1 within the pH range. Assuming the second reaction (*M2*) as the RLS, we found two possible pH independent reaction orders, $\rho_{M2\omega}$ = 1 for high overpotential (labeled ω) and $\rho_{M2\beta}$ = 2 for low overpotential (labeled β). At intermediate overpotentials, the reaction order was fractional and depended on pH. This is again analogous to the observations of the Tafel slopes with pH (Fig. 4b). Moving to *M3* as the RLS, the reaction order at high overpotentials (labeled ω) retained the previous value of $\rho_{M3\omega}$ = $\rho_{M2\omega}$ = 1, while the reaction order at low overpotential was $\rho_{M3\beta}$ = 3, both independent of pH. An intermediate reaction order of $\rho_{M3\delta}$ = 2 is expected but barely reached at pH 14 (Fig. 5c). Instead, the reaction order was fractional and depended on the pH. The sigmoidal shape of the trend and comparison with Fig. 5d (i.e., a different set of kinetic constants) suggested that $\rho_{M3\delta}$ = $\rho_{M3\beta}$ = 3 at low pH. For the later RLS of *M4* and *M5* (Fig. 5e,f), the reaction order at low overpotential ($\rho_{Mi\beta}$) increased with the number of transferred OH⁻ (up to 4 for *M5*), while the reaction order depended on pH for higher overpotentials. According to our literature survey, reaction orders $\rho$ > 2 have not been reported yet. Mainly one can find reaction orders of either $\rho$ = 1 [19,20,56,60] or $\rho$ = 2 [61–64] or fractional values below 2, such as $\rho$ = 1.5 [56] for first row transition metal oxides in alkaline electrolytes. This hints at an early RLS in experiments but cannot exclude a later RLS where the expected higher reaction order was not resolved.

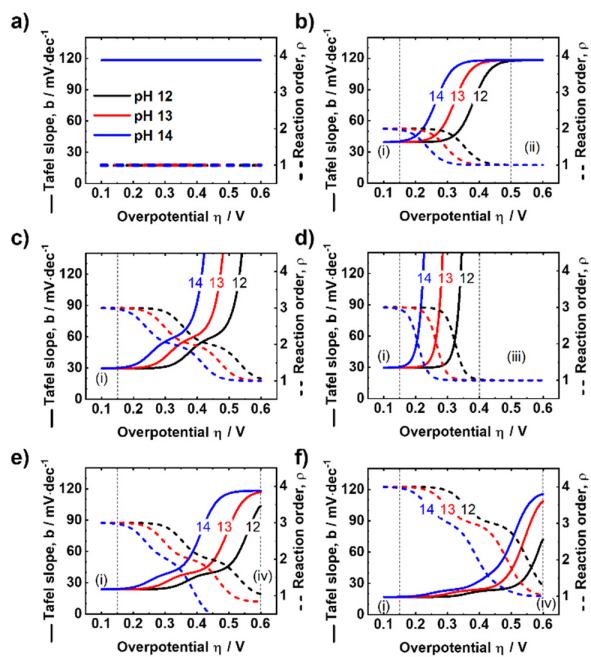

**Fig. 6**. Simulated Tafel slopes (left y-axis) and reaction order (right y-axis) as function of overpotential (E-E$^0$) for pH 12, 13 and 14 the RLS of (a) *M1*, (b) *M2*, (c) *M3* (d) *M3'*, (e) *M4* (e) and (f) *M5* steps. *M3* and *M3'* have the same RLS but were simulated with different sets of kinetic constants found in Table 1. $\alpha$-$\omega$ indicate potential regions with constant mechanistic parameters. Dataset in ref.[54].

## *Synergy between Tafel slope and reaction order*

In order to investigate the complementary of the Tafel slope and reaction order, we plotted simulated Tafel slopes overlaid with the corresponding reaction orders as function of overpotential for pH 12, 13 and 14 (Fig.6). For *M1* as the RLS, both Tafel slope and reaction order were constant as expected from the above discussions (Fig. 6a). When *M2* was the RLS, there were two overpotential regions labeled (i) and (ii) at the lowest and highest overpotentials, respectively, for which Tafel slope and reaction order were simultaneously constant (Fig. 6b). These are the overpotential ranges that can be meaningfully evaluated for mechanistic insight and to benchmark mechanisms. For intermediate overpotentials, both Tafel slope and reaction order changed as function of pH. When *M3* is the RLS, three different combinations of Tafel slope and reaction order can be expected but, only one region of constant values (region (i) in Fig. 6c) was found at low overpotentials in our simulations where a second region is just outside the highest investigated overpotentials. Additionally, there was a very narrow overpotential region with the third combination which could not be clearly distinguished from the transition region with overpotential- and pH dependence. With the alternative set of kinetic constants for the same RLS, the intermediate combination vanished and the high overpotential Tafel slope and reaction order were constant (region (iii) in Fig. 6d). *M4* and *M5* as the RLS are similar to the case described for *M3* where only the low overpotential region (i) gives access to constant Tafel slopes and reaction orders (Fig. 6e,f). It should be noted that the reaction order becomes zero for an electrochemical RLS after a chemical RLS at high overpotential and pH as can be seen in Fig. 6e. In summary, the most insightful region for mechanistic analysis can be found at low overpotential. Secondary information can be gained at high overpotential, particularly regarding whether the RLS is chemical or electrochemical. Experimentally, the product current of the evolved oxygen should be detected for reliable determination of the Tafel slope and reaction order, particularly at low overpotential where they may be in competition with metal redox peaks [65]. The high overpotential region may be inaccessible due to excessive bubble formation.

We now summarize our insights in a schematic catalytic cycle for the adsorbate mechanism (Fig. 7), where unique combinations of the Tafel slope and reaction order are highlighted. These unique combinations are found at low overpotential and pH, while the secondary combinations can be identical to previous steps of the same type if resolved, e.g., *M2* (and all other steps with an electrochemical RLS) behaved like *M1* at high overpotential, i.e., they exhibited the combination of $\rho$ = 1 and *b* = 118 mV/dec. It means that only the last $OH^-$ transfer matters and all previous steps can be neglected in the calculation of the reaction order as proposed elsewhere [20]. The only chemically limiting step *M3* has distinct Tafel slopes from the electrochemically limited steps. Overall, we hope that our analysis and in particular the unambiguous assignments in Fig. 7 provide a benchmark for the identification of this mechanism in experimental studies.

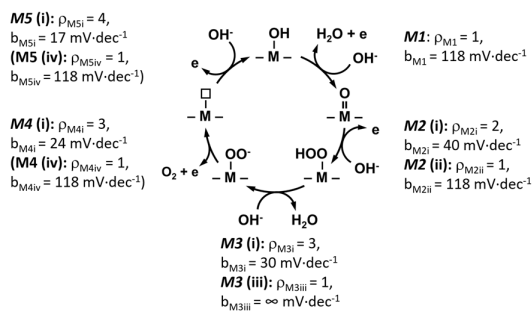

**Fig. 7**. Catalytic cycle for adsorbate mechanism with corresponding unique combinations of the Tafel slope and reaction order.

## Conclusions

Two key mechanistic parameters were calculated using a microkinetic approach. We made our simulations for the commonly discussed adsorbate mechanism [11,12,24]. A hydroxylated surface was considered as the starting point in contrast to DFT and other microkinetic works based on empty sites [10].The simulated Tafel plots showed an increase of possible Tafel slopes with the number of preceding steps. Not all Tafel slopes may be observed depending on the kinetic constants, which in turn affected the population of intermediate surface states. If the surface coverage of a specific intermediate was low, then the corresponding Tafel slope might not be observed. When the surface coverage was not constant, the Tafel slope strongly depended on applied overpotential and pH, which resulted in regions with no well-defined values of the slope. We further investigated the effect of the transfer coefficient on the value of the Tafel slope, which had its largest influence at high overpotential. We demonstrated that the RLS cannot be identified by a single Tafel slope. Instead, an additional mechanistic parameter was required to distinguish between possible rate-limiting steps and hereafter mechanisms. Complementary to Tafel slopes, we provided simulations of the reaction order with respect to pH. We showed that this parameter is not affected by the transfer coefficient. We discussed that the most insightful information can be obtained from the low overpotential region, where the influence of the (often unknown) transfer coefficient was lowest and potential- as well as pH-independent values of both the Tafel slope and reaction order were observed. Thus, knowing the maximum value of the reaction order accompanying by the Tafel slope at low overpotential made it possible to distinguish the RLS of the adsorbate mechanism. The main challenge is stabilizing those RLS over a large potential range to obtain small Tafel slope values experimentally. Nonetheless, our work provides clear guidelines to experimentalists for the identification of the RLS in the adsorbate mechanism using the observed values of the Tafel slope and reaction order.

**Acknowledgements**

This project has received funding from the European Research Council (ERC) under the European Union's Horizon 2020 research and innovation programme under grant agreement No. 804092.

**Conflict of Interest**

We declare no conflict of interest

**Data statement**

The data that support the findings of this study are openly available in Figshare at https://doi.org/10.6084/m9.figshare.17122220

Supporting Information

**Calculation of the Tafel slope and reaction order of the oxygen evolution reaction in alkaline electrolytes for the adsorbate mechanism**

Denis Antipin, Marcel Risch*

Nachwuchsgruppe Gestaltung des Sauerstoffentwicklungsmechanismus, Helmholtz-Zentrum Berlin für Materialien und Energie GmbH, Hahn-Meitner Platz 1, 14109 Berlin, Germany

* marcel.risch@helmholtz-berlin.de

9 pages, 6 supporting figures, 2 supporting tables

**Example of the Tafel slope calculation**

Assume first two steps (M1 and M2) of adsorbate mechanism (calculations for further steps were made similarly to them):

  M1  M-OH + OH$^-$ ⇌ M=O + H$_2$O + e

  M2  M=O + OH$^-$ ⇌ M-OOH + e

Current density for the first reaction (M1) can be written as:

$$j_1 = I/A = n \cdot F \cdot r_1 \tag{Eq. S1}$$

The reaction rate is equal to:

$$r_1 = k_{+1} \cdot a_{OH} \cdot \theta_{OH} \tag{Eq. S2}$$

As is was mentioned in the main text, we assumed surface coverage equal to $\theta_{OH}$ = 1 for the first step in the mechanism. For Tafel analysis a hydroxide activity of $a_{OH}$ = 1 was used.

Since the we first reaction is electrochemical, the expression for the kinetic constant has a form:

$$k_{+1} = k^0_{+1} \cdot e^{(1-\alpha) \cdot f \cdot \eta} \tag{Eq. S3}$$

Combining Eqs. S1-S3, we formulate the final expression for the current density for the first forward step:

$$j_1 = n \cdot F \cdot k^0_{+1} \cdot e^{(1-\alpha) \cdot f \cdot \eta} \cdot a_{OH} \tag{Eq. S4}$$

Using values for k$^0$$_{+1}$ of 1 from Table 1, we could plot voltammogram for the later Tafel analysis (Fig. 2a of main text).

If we consider second reaction (M2) as a RLS, an assumption of pre-equilibrated previous steps was used. It means, that the reaction rates of the forward and backward steps prior to RLS are equal:

$$r_1 = r_{-1} \tag{Eq. S5}$$

The reaction rate of the backward step $r_{-1}$ can be written as:

$$r_{-1} = k_{-1} \cdot a_{H_2O} \cdot \theta_O \tag{Eq. S6}$$

The expression for the kinetic constant has a form:

$$k_{-1} = k^0_{-1} \cdot e^{-\alpha \cdot f \cdot \eta} \tag{Eq. S7}$$

Combining Eqs. S6-S7, the final expression for the reaction rate of the backward step can be written:

$$r_{-1} = k^0_{-1} \cdot e^{-\alpha \cdot f \cdot \eta} \cdot a_{H_2O} \cdot \theta_O \tag{Eq. S8}$$

After a substitution of Eq. S2 and Eq. S8 in equality Eq. S5, we got first equation for surface coverages:

$$k_{+1} \cdot a_{OH} \cdot \theta_{OH} = k^0_{-1} \cdot e^{-\alpha \cdot f \cdot \eta} \cdot a_{H_2O} \cdot \theta_O \tag{Eq. S9}$$

Second equation came from a fact that overall surface coverage is equal to 1. For the first reaction M1 it has the following form:

$$\theta_{OH} + \theta_O = 1 \tag{Eq. S10}$$

Solving system of two equations Eq. S9 and Eq. 10 with respect to $\theta_{OH}$ and $\theta_O$, final expression for the surface coverage were obtained:

$$\theta_{OH} = \frac{a_{H_2O}}{a_{H_2O} + \frac{k^0_1}{k^0_{-1}} exp(f*\eta)*a_{OH}} \tag{Eq. S11}$$



$$\theta_O = \frac{\frac{k_1^0}{k_{-1}^0} exp(f*\eta)*a_{OH}}{a_{H_2O}+\frac{k_1^0}{k_{-1}^0} exp(f*\eta)*a_{OH}} \tag{Eq. S12}$$

Akin to the first reaction, the current density and the reaction rate for the second one can be written as:

$$j_2 = I/A = n \cdot F \cdot r_2 \tag{Eq. S13}$$

$$r_2 = k_{+2} \cdot a_{OH} \cdot \theta_O \tag{Eq. S14}$$

After combining Eqs. S10, S13-14, we got the final expression for the current density:

$$j_2 = n*F*k_{+2}^0*a_{OH}*\frac{\frac{k_1^0}{k_{-1}^0} exp(f*\eta)*a_{OH}}{a_{H_2O}+\frac{k_1^0}{k_{-1}^0} exp(f*\eta)*a_{OH}}*e^{(1-\alpha)*f*\eta} \tag{Eq. S15}$$

Using values for kinetic constants from Table 1, we could plot voltammogram for the later Tafel analysis (Fig. 2b of main text).

**Figures**

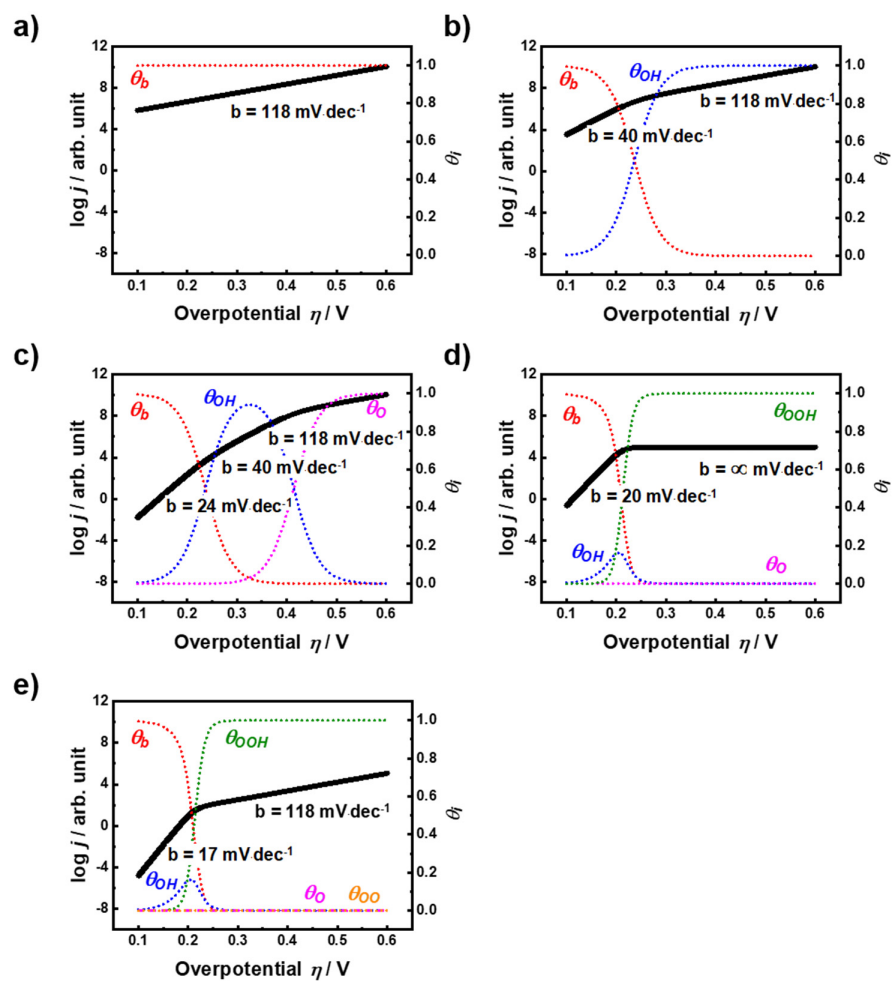

**Fig. S1.** Simulated Tafel plots assuming M5 as the initial step. Same set of kinetic constants (Table 1 of the main text) used for this Tafel plot, except $k_{-4}$ ($10^6$ instead of $10^{12}$, see Table S1).

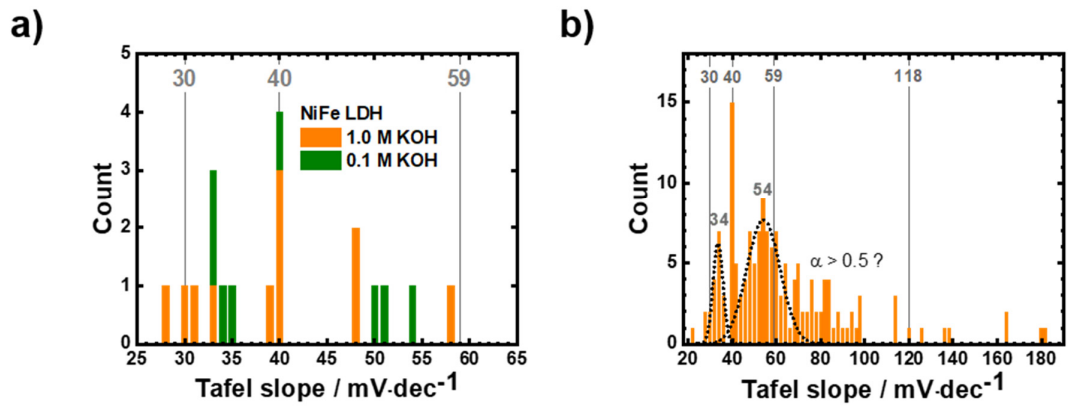

**Fig. S2.** (a) Histogram of Tafel slopes reported for NiFe LDH in 1.0 M and 0.1 M KOH. Binning width was 1 mV/dec. Data taken from ref. [1]. (b) Histogram of Tafel slopes reported for various electrocatalyst. Data taken from [1-5]. Binning was 5 mV/dec. The grey lines indicate common Tafel slopes assuming $\alpha$ =0.5.

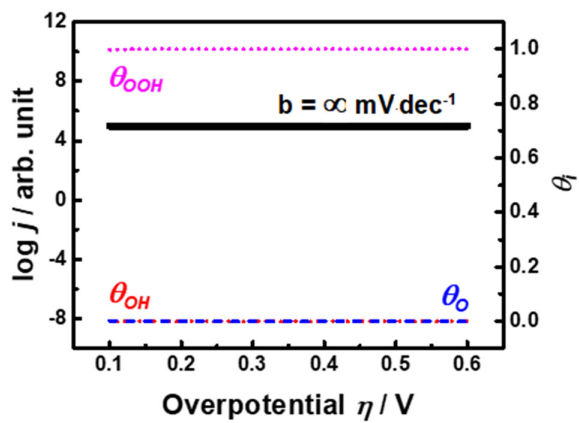

**Fig. S3.** Simulated Tafel plot assuming M3" step is RLS. Different set of kinetic constants (Table S2) used for this Tafel plot results in one dominant surface coverage and one Tafel slope.

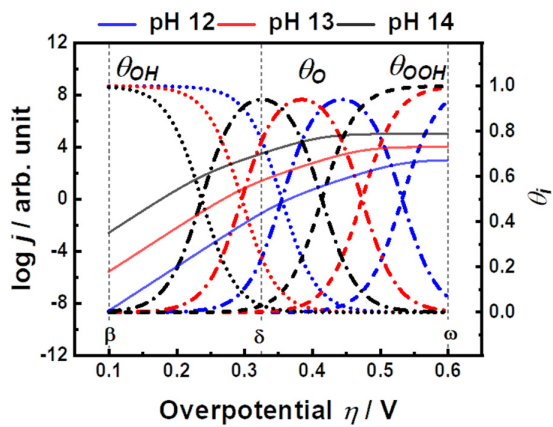

**Fig. S4.** Simulated Tafel plots between pH 12 and 14 assuming *M3* as the RLS. The kinetic constants may be found in Table 1. All simulations use a transfer coefficient of $\alpha$ = 0.5. $\beta$, $\delta$ and $\omega$ denote potentials of interest that are discussed in the main text.

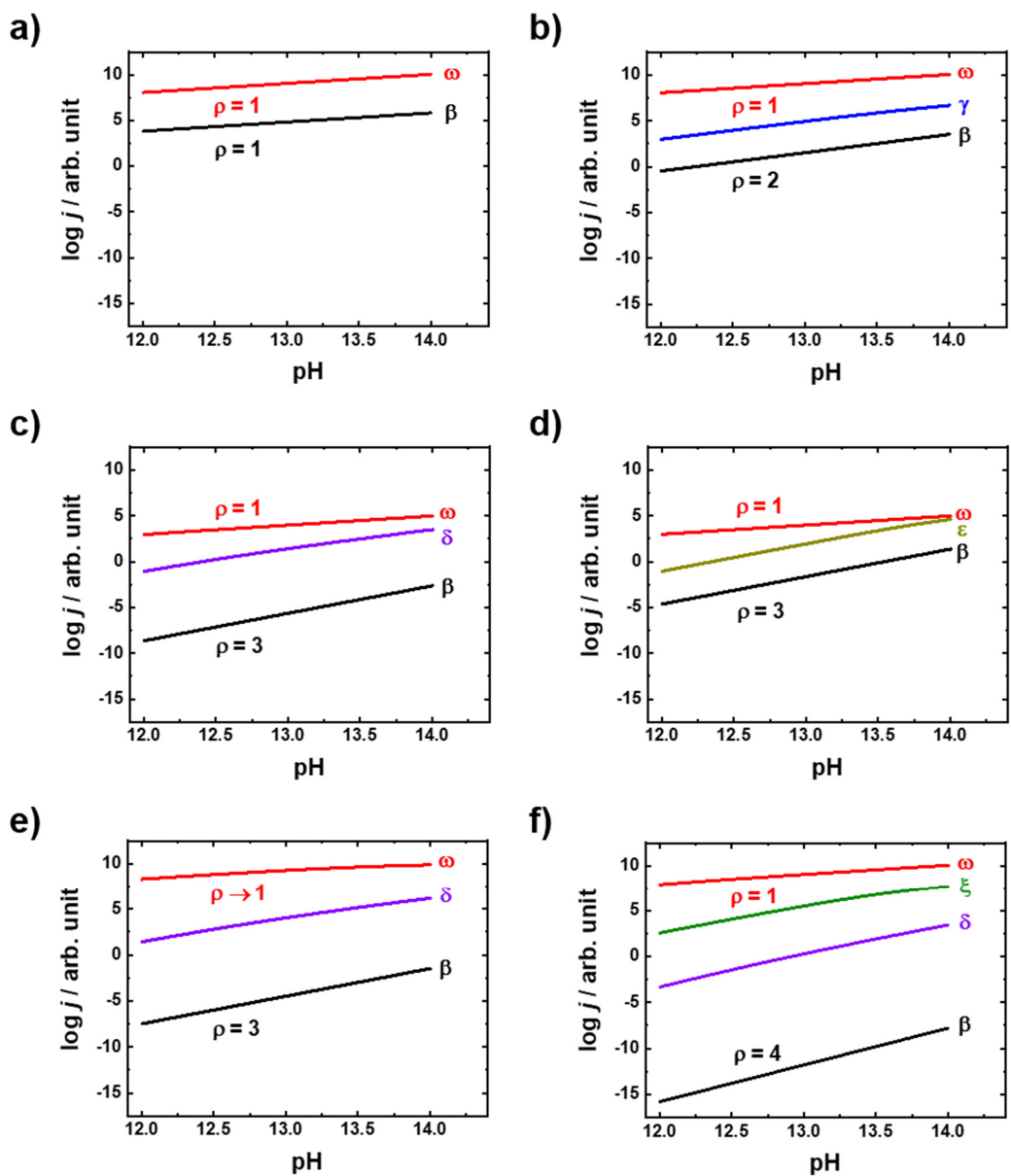

**Fig. S5.** Simulated current densities as function of pH at key potentials identified in Fig. 2 assuming the RLS of (a) *M1*, (b) *M2*, (c) *M3*, (d) *M3'*, (e) *M4* and (f) *M5*. *M3* and *M3'* have the same RLS but were simulated with different sets of kinetic constants found in Table 1. All simulations use a transfer coefficient of $\alpha = 0.5$.

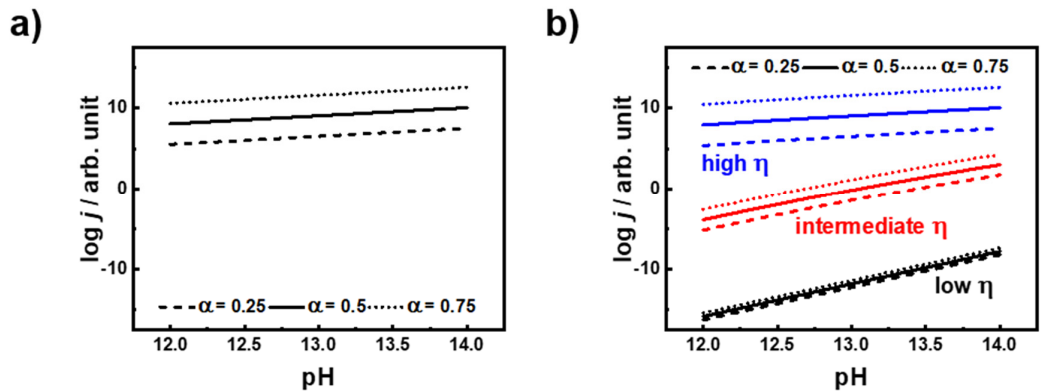

**Fig. S6.** Simulated log *j* vs. pH plots assuming the RLS of (a) *M1* and (b) *M5* steps with different values of the transfer coefficient.

## Tables

**Table S1.** Values of kinetic constants used in the simulation for Fig. S1. The same mechanism with different step order (M5 – M4) was used for the simulations.

| Kinetic constant | Rate-limiting step | | | | |
|---|---|---|---|---|---|
| | *M5* | *M1* | *M2* | *M3* | *M4* |
| $k_{+1}$ | 1 | $10^3$ | $10^3$ | $10^3$ | $10^3$ |
| $k_{-1}$ | – | $10^7$ | $10^7$ | $10^7$ | $10^7$ |
| $k_{+2}$ | – | 1 | $10^3$ | $10^3$ | $10^3$ |
| $k_{-2}$ | – | – | $10^{10}$ | $10^{10}$ | $10^{10}$ |
| $k_{+3}$ | – | – | 1 | $2*10^{12}$ | $2*10^{12}$ |
| $k_{-3}$ | – | – | – | $10^{12}$ | $10^{12}$ |
| $k_{+4}$ | – | – | – | 1 | 10 |
| $k_{-4}$ | – | – | – | – | $10^6$ |
| $k_{+5}$ | – | – | – | – | 1 |
| $k_{-5}$ | – | – | – | – | – |

**Table S2.** Values of kinetic constants used in the simulation for Fig. S3. *M3* and *M3"* are the same rate-limiting step with different sets of constants

| Kinetic constant | Rate-limiting step | |
|---|---|---|
| | *M3* | *M3"* |
| $k_{+1}$ | $10^3$ | $10^3$ |
| $k_{-1}$ | $10^7$ | $10^2$ |
| $k_{+2}$ | $10^3$ | $10^3$ |
| $k_{-2}$ | $10^{10}$ | $10^2$ |
| $k_{+3}$ | 1 | 1 |
| $k_{-3}$ | – | – |